\documentclass[10pt,twocolumn,english,aps,prl,showpacs,superscriptaddress,groupaddress,floatfix,graphics,graphicx]{revtex4-2}

\usepackage[T1]{fontenc}
\usepackage{color}
\usepackage[english]{babel}
\usepackage{float}
\usepackage{amsmath}
\usepackage{amssymb}
\usepackage{graphicx}
\usepackage{xcolor}
\usepackage{soul}
\usepackage[bookmarks]{hyperref}
\hypersetup{colorlinks,linkcolor=blue,citecolor=blue,urlcolor=blue}
\usepackage[all]{hypcap}

\usepackage{soul}
\usepackage{comment}
\usepackage{orcidlink}

\usepackage{amsmath}
\usepackage{bm}
\usepackage{siunitx}
\usepackage{nicefrac}

\usepackage{graphicx}
\usepackage{xcolor}
\usepackage{xspace}
\usepackage{float}

\newcommand{\ket}[1]{| {#1} \rangle} 
\newcommand{\bra}[1]{\langle {#1} |} 
\DeclareMathAlphabet\mathbfcal{OMS}{cmsy}{b}{n}

\usepackage{hyperref}
\newcommand{\SAVBA}{\affiliation{Institute of Informatics, Slovak Academy of Sciences, 84507 Bratislava, Slovakia}}
\newcommand{\FFBG}{\affiliation {Faculty of Physics, University of Belgrade, 11001 Belgrade, Serbia}}
\newcommand{\UPJS}{\affiliation{Institute of Physics, Pavol Jozef \v{S}af\'{a}rik University in Ko\v{s}ice, 04001 Ko\v{s}ice, Slovakia}}
\newcommand{\SAVKE}{\affiliation{Institute of Experimental Physics, Slovak Academy of Sciences, 04001 Ko\v{s}ice, Slovakia}}

\newcommand{\NTC}{\affiliation{New Technologies Research Centre, University of West Bohemia, CZ-301 00 Pilsen, Czech Republic}}

\begin{document}
\title{
Stacking switching between correlation-protected radial Rashba field and persistent spin textures in graphene encapsulated by 1T-TaS$_2$ monolayers
}

\author{Juraj Mnich}\email[Corresponding author: ]{juraj.mnich@student.upjs.sk} \UPJS 
\author{Marko Milivojevi{\' c}} \SAVBA \FFBG
\author{Martin Gmitra} \UPJS \SAVKE \NTC
\date{\today}

\begin{abstract}
We investigate the electronic structure, spin textures, and charge to spin/orbital transport in graphene encapsulated by 1T-TaS$_{2}$ monolayers in the charge density wave phase. Using first-principles calculations, tight-binding modeling, and the Kubo formalism, we show that the encapsulation stacking dictates fundamentally distinct transport regimes. In the asymmetrical (AA) stacking, proximity fields from both interfaces constructively interfere, yielding a cumulative Rashba phase of nearly $\pi/2$. This pure radial Rashba spin pattern leads to the unconventional Rashba-Edelstein effect, which robustly dominates over the conventional response by a factor of 35 across a wide energy range. Conversely, the symmetrical (AA') stacking preserves a horizontal mirror symmetry, establishing a stable, purely out-of-plane persistent spin texture. Furthermore, the computed orbital Hall effect is exceptionally efficient, surpassing the spin Hall effect by three orders of magnitude. Within the proximity-induced spectral gaps, the orbital Hall conductivity exhibits a finite plateau, whereas the spin Hall conductivity vanishes. Our findings establish graphene encapsulated heterostructures as a promising system for realizing distinct charge to spin and charge to orbital interconversion regimes determined by the choice of stacking order.
\end{abstract}
\maketitle

\section{Introduction}
Van der Waals (vdW) heterostructures~\cite{Geim2013,Wang2021} offer a versatile platform for engineering material properties utilizing the proximity effect. In the field of spintronics~\cite{ZFS04,Han2014,Sierra2021,Kurebayashi2022,Zollner2025:NRP}, 
graphene-based vdW heterostructures have been recognized as potential platforms for spintronic and orbitronic applications~\cite{Gmitra2015,Gmitra2016,Gmitra2017,Sierra2021,Cysne2025:npjSpin}, due to the demonstrated possibility to independently induce and control spin-dependent interactions in graphene~\cite{Avsar2014, Zollner2016, Zollner2021, ZJN+23}. 
To further modify spin textures in graphene, different approaches have been employed, such as electric field~\cite{ZJN+23}, substrate~\cite{MGK+24}, vertical pressure~\cite{FMZ+21}, or twist angle modulation~\cite{David2019,Naimer2021,Peterfalvi2022,Veneri2022,Lee2022b}. 

It is known that the Peierls instability distorts a periodic lattice \cite{Pasquier2019} and triggers low-temperature charge density wave (CDW) phases in some transition-metal dichalcogenides (TMDCs). The most prominent examples are TaS$_2$~\cite{Wilson1975,Brouwer1980,Miller2018}, TaSe$_2$~\cite{Miller2018,Zhang2020,Jiang2021}, and TaTe$_2$~\cite{Miller2018,Jiang2021}, or NbS$_2$~\cite{Tresca2019} and NbSe$_2$~\cite{Calandra2018,Liu2021,Liu2021b}. Recently, using first-principles calculations and effective tight-binding modeling, the graphene/1T-TaS$_2$~\cite{SMK+23} heterostructure has been proposed as an all-in-one device in which the temperature can be used as a control knob to tune between different correlated phases of 1T-TaS$_2$, affecting the proximity-induced interactions in graphene. Similarly to what is observed in twisted structures, the presence of the CDW phase enhances the radial component of the spin textures with the Rashba phase $\phi_{\rm R} \approx \pi/4$, but without physical twisting. This manifestation introduces a paradigm of twistronics-without-a-twist, where electronic correlations and the complex lattice distortions of the CDW phase act as a direct proxy for structural twist-angle engineering \cite{Veneri2022}. By using the distinct phase transitions of $\text{1T-TaS}_2$ as an external control knob rather than mechanical rotation, one can manipulate the spin texture and reconfigure the interface symmetries. Furthermore, it was experimentally proven that the same heterostructure offers the possibility of omnidirectional charge to spin interconversion~\cite{Chi2024}, providing an experimentally verifiable fingerprint. 

\begin{figure}[h!]
    \centering
    \includegraphics[width=0.98\columnwidth]{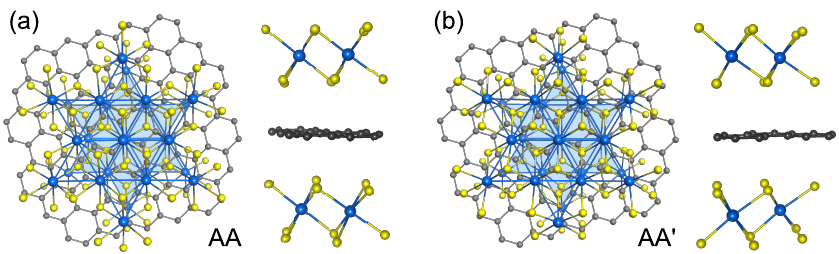}
    \caption{Structural model of encapsulated graphene between 1T-TaS$_2$ monolayers in the CDW phase of the David star periodic lattice distortions.
    (a)~asymmetric AA stacking of the 1T-TaS$_2$ monolayers in the perspective top-like and side view, and
    (b)~symmetric AA' stacking with a horizontal mirror plane.
    The blue and yellow balls represent the Ta atoms and the S atoms, and the gray balls are the C atoms.
    }
    \label{fig:structure}
\end{figure}

Here, we extend this concept by stacking control over the spin texture in a 1T-TaS$_2$/graphene/1T-TaS$_2$ heterostructure, as shown in Fig.~\ref{fig:structure}. We study the effect of stacking, encapsulating graphene between two 1T-TaS$_2$ monolayers in the CDW phase. Combining the two 1T-TaS$_2$ layers in the asymmetrical (AA) stacking configuration leads to an enhancement of the Rashba phase to the radial limit ($\phi_{\rm R} \approx \pi/2$) due to constructive interference from both sides of the proximitized graphene. 
On the other hand, by flipping one of the 1T-TaS$_2$ layers to form the symmetrical (AA') stacking configuration, the system preserves a horizontal mirror plane symmetry that effectively forces a permanent out-of-plane spin texture. 
The spin textures of these encapsulated graphene heterostructures are remarkably robust against electric-field modulations, anchoring the system within the systems with a pure radial Rashba phase regime \cite{Frank2024,Kang2024} and persistent spin textures protected by symmetry \cite{Przybysz2026:APL}.
Ultimately, engineering these proximity-induced interactions aims to optimize spin-charge interconversion mechanisms serving as the operational foundation for spintronics.

Crucially, the spin degree of freedom is generally accompanied by the orbital degree of freedom. The transverse generation of a charge-neutral stream of orbital angular momentum driven by a longitudinal electric field is known as the orbital Hall effect (OHE) \cite{Bernevig2005:PRL,GJK18}. This phenomenon operates similarly to the spin Hall effect (SHE) \cite{Dyakonov1971:PLA,Sinova2015:RMP}, where an electric field drives a perpendicular spin current, an effect that has become a cornerstone of modern spintronics for spin current generation and detection schemes. While the intrinsic SHE is fundamentally tied to a finite SOC or exchange coupling, the OHE can develop entirely independently of relativistic effects, subsequently acting as a source that can be converted into a spin current via spin-orbit interactions \cite{GJK18}. Moreover, both intrinsic SHE and OHE can emerge in centrosymmetric materials through momentum-space orbital textures, which are inherently ubiquitous across crystal structures regardless of inversion symmetry, contrasting sharply with the more symmetry-restricted spin textures \cite{GJK18, SBB24}. Within the modern theory of orbital magnetization based on the mixed Berry curvature \cite{TCV05}, a finite OHE can be achieved even within effective tight-binding frameworks with orbital basis carrying an intrinsically vanishing atomic orbital angular momentum \cite{BMG23}. Exploiting this feature, Bhowal and Vignale \cite{BV21} demonstrated that the OHE provides a robust alternative to the conventional valley Hall effect in gapped graphene architectures, where the underlying orbital Berry curvature enables the manifestation of a generalized valley orbital Hall response. In this work, we show that the proximity-induced effects in encapsulated graphene lead to a giant OHE -- three orders of magnitude larger than the SHE.

This paper is organized as follows. In Section~\ref{First-calculations}, we describe the first-principles calculations details, whereas in Section~\ref{Bandstructure} the electronic band structure of the AA and AA' 1T-TaS$_2$/graphene/1T-TaS$_2$ heterostructures is analyzed. 
The electronic structure of graphene close to the Dirac cone can be described using a simple model given in Section~\ref{In-plane}.
Using the model, the spin and orbital angular momentum accumulation is presented in Sec.~\ref{sec:charge-spin}, and SHE and OHE are discussed in \ref{sec:orbital-cond}.
In Sec.~\ref{sec:conclusions}, we present the final remarks.

\section{Results}
\subsection{First-principles calculation details}\label{First-calculations}
We model the encapsulated trilayer system using a commensurate supercell consisting of a $5\times 5$ graphene sheet matched to a $\sqrt{13}\times\sqrt{13}\text{ R}13.9^{\circ}$ superstructure of the $\text{1T-TaS}_2$ monolayers. To achieve commensurability, the graphene lattice is compressively strained by approximately $1.4\%$ to match the experimental $\text{1T-TaS}_2$ CDW lattice constant of $12.1323\text{ \AA}$~\cite{Spijkerman1997}. This low-temperature commensurate CDW phase manifests as a periodic lattice distortion characterized by a clustering of the Ta atoms into a David star pattern. This atomic rearrangement structurally reconstructs the pristine $1\times1$ monolayer into an enlarged supercell that is rotated by $13.9^{\circ}$ with respect to the original primitive lattice vectors~\cite{Spijkerman1997,Stahl2020}.
The electronic structure calculations were performed using the {\sc Quantum Espresso} suite~\cite{Giannozzi2009,Giannozzi2017}, which utilizes a plane wave basis for DFT calculations~\cite{Hohenberg1964}. All the calculations were performed using the Perdew-Burke-Ernzerhof exchange-correlation functional~\cite{Perdew1996} for the projector augmented wave method~\cite{Kresse1999}. In all studied configurations, we have assumed a vacuum of approximately 15~\AA~in the $z$-direction (direction perpendicular to the heterostructure) to avoid the interaction between the periodic images. The relaxation of atomic positions was performed using the spin-unpolarized DFT calculations, with a threshold force of $10^{-4}$ Ry per Bohr radius, where to properly describe the interlayer distance between the heterostructure constituents, the semiempirical Grimme-D2 van der Waals corrections were included~\cite{Grimme2006,Barone2009}.
The non-collinear DFT calculations including SOC were performed using the fully relativistic pseudopotentials~\cite{DalCorso2014}. Also, the dipole correction~\cite{Bengtsson1999} was applied to properly determine the Dirac point energy offset due to dipole electric field effects. Finally, small Methfessel-Paxton energy level smearing~\cite{Methfessel1989} of 1~mRy was used in all the studied cases,  alongside the kinetic energy cut-off of 53~Ry and the $9\times 9$ mesh of $k$-points for sampling the first Brillouin zone. 

\subsection{Model Hamiltonian}\label{In-plane}
Low-energy electronic states of the encapsulated graphene 
can be described by the effective model Hamiltonian containing $\pi$-bands~\cite{Kochan2012,Kochan2017}. 
In the vicinity of the $K/K'$ point $(\kappa=1/-1)$ the Hamiltonian can be expressed as
\begin{align}\label{C3Model}
H_{\kappa}&=E_0\sigma_0\otimes S_0+\Delta\sigma_z \otimes S_0\nonumber\\
&+\hbar v_{\rm F}(\kappa k_x\sigma_x+k_y\sigma_y)\otimes S_0\nonumber\\
& -\lambda_{\rm R}U^\dagger(\kappa \sigma_x\otimes S_y-\sigma_y\otimes S_x)U\nonumber\\
&+\kappa\left(\lambda_{\rm I}^{\rm{A}}\frac{\sigma_z+\sigma_0}2+ \lambda_{\rm I}^{\rm{B}}\frac{\sigma_z-\sigma_0}2\right)\otimes S_z,
\end{align}
where $E_0$ represents the Dirac point offset with respect to the heterostructure's Fermi energy, $\Delta$ is the sublattice-dependent staggered potential, $v_{\rm{F}}$ is the Fermi velocity, and $\lambda_{\mathrm I}^{\mathrm A}$ and $\lambda_{\mathrm I}^{\mathrm B}$ are the sublattice-resolved intrinsic SOC parameters.
We formally distinguish between the Pauli spin matrices $S_{0,x,y,z}$ acting in a spin space and Pauli matrices $\sigma_{0,x,y,z}$ acting in a sublattice space (A, B), 
The AA stacking possesses the ${\bf C}_{3}$ symmetry with a general Rashba form, where $U=\sigma_{0}\otimes e^{\mathrm{i} S_z \phi_{\mathrm R}/2}$ is the the unitary operator, and Rashba phase $\phi_{\rm{R}}$ depends on the SOC matrix elements between $d$-orbitals of metal atoms in TMDC \cite{David2019} and the tunneling matrix elements between the graphene and TMDC atoms \cite{Peterfalvi2022}.
For the ${\bf C}_{3\rm{h}}$ symmetry of the AA' stacking, the Rashba term is absent.

\subsection{Electronic band structure}\label{Bandstructure}
\begin{figure}[h]
\centering
\includegraphics[width=0.98\columnwidth]{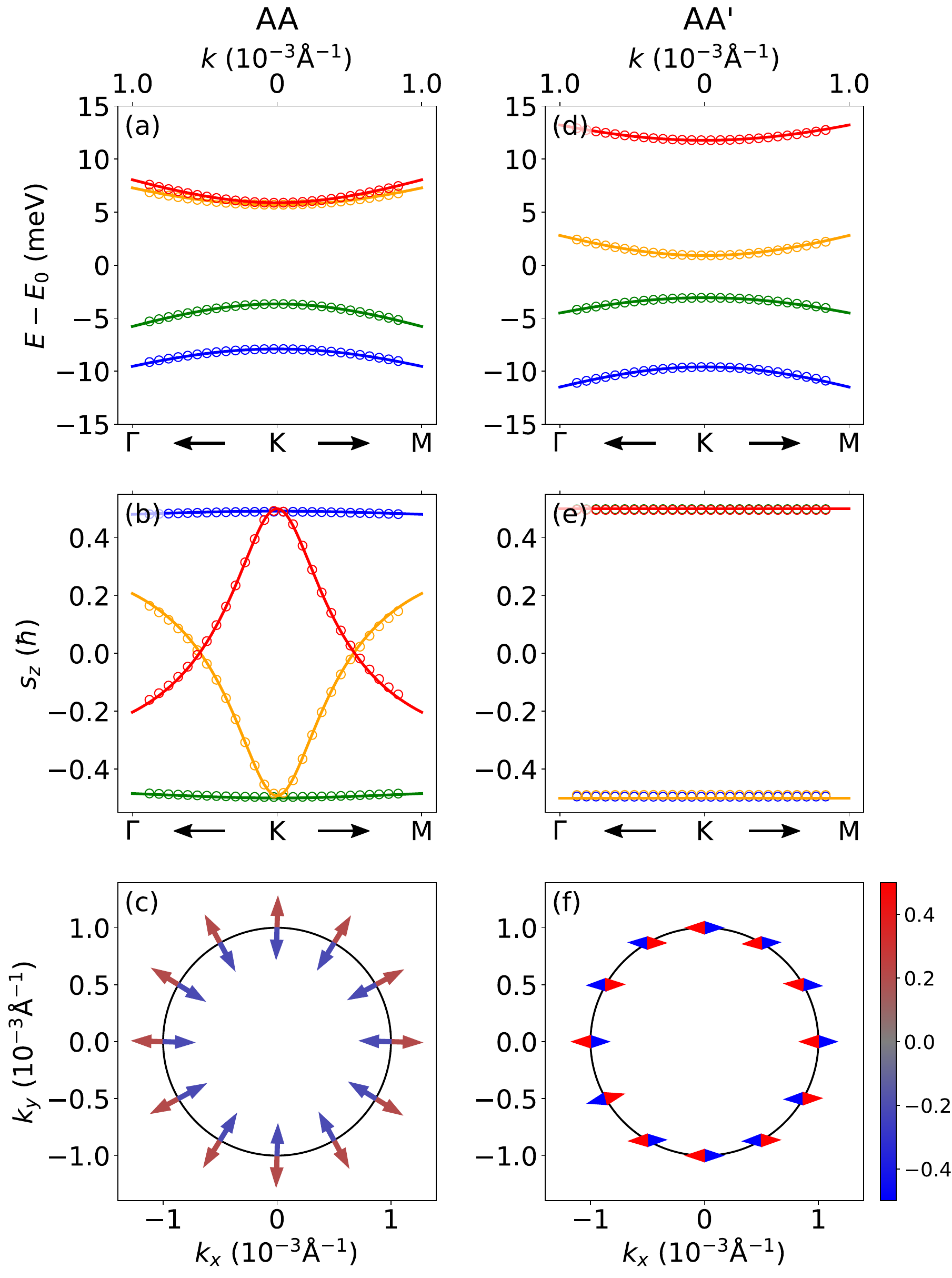}
\caption{Low-energy electronic band structure and spin expectation values of encapsulated graphene between 1T-TaS$_2$ monolayers in the CDW phase.
Left panels (a)–(c) correspond to the asymmetric AA stacking configuration, and right panels (d)–(f) denote the symmetric AA' stacking configuration.
(a), (d) Energy bands near the charge neutrality point, where colored circles represent DFT calculations and solid lines indicate the effective tight-binding model fits. 
The color coding tracks band indexing: red/orange and green/blue denote the conduction-like and valence-like bands, respectively. 
(b), (e) Out-of-plane spin expectation value $s_z$ along the high-symmetry $k$-path. 
(c), (f) Spin textures evaluated for the conduction-like bands along a circular contour centered at the $K$ point with a radius of $|k| = 10^{-3}$~\AA$^{-1}$. In the bottom row, the color scale indicates the magnitude and sign of the out-of-plane component $s_z$ (blue and red), while arrows depict the in-plane spin projections. For the AA' stacking (f), the in-plane components vanish identically due to horizontal mirror symmetry.}
    \label{fig:DFT_fit}
\end{figure}

We investigate two stacking configurations of graphene encapsulated by two 1T-TaS$_2$ monolayers in the CDW phase: the asymmetrical (AA) stacking and the symmetrical (AA') stacking with horizontal mirror plane symmetry. In both cases, we consider only the top stacking where the center of the David star of 1T-TaS$_{2}$ is above/below the carbon atom of graphene as depicted in Fig.~\ref{fig:structure}. The comparison of band structures and spin expectation values near the Dirac point of the DFT calculation and tight-binding fits is shown in Fig.~\ref{fig:DFT_fit}, where the tight-binding model faithfully reproduces both the electronic band structure and spin texture using the parameters listed in Tab.~\ref{tab:TBparam}.
\begin{table}[h!]
\caption{\label{tab:TBparam} Tight-binding model parameters of graphene encapsulated by AA and AA' stacked 1T-TaS$_2$ monolayers in CDW phase.} 
\centering
\small
\setlength{\tabcolsep}{1.5em}
\begin{tabular}{l|rr}\hline
E (V/nm)& AA & AA' \\\hline\hline
$v_{\rm{F}}$ ($10^6$ m/s) & 0.758 & 0.737\\
$E_0$ (meV) & 307.56 & 306.02\\
$\Delta$ (meV) & 5.73 & 6.33\\
$\lambda_{\mathrm {I}}^{\rm{A}}$ (meV) & 0.15 & 5.43\\
$\lambda_{\mathrm {I}}^{\rm{B}}$ (meV) & 2.07 & -3.26\\
$\lambda_{\mathrm R}$ (meV) & -0.61 & 0\\
$\phi_{\mathrm R}$ (deg) & 88.35 & 0 \\\hline
\end{tabular}
\end{table}

As was already shown in \cite{SMK+23}, the breaking of horizontal mirror plane symmetry of graphene by a single layer of 1T-TaS$_{2}$ causes the emergence of a Rashba phase of about $45^{\circ}$. Placing another copy of 1T-TaS$_{2}$ from the opposite side of the graphene layer builds the AA stacking. In this configuration, the graphene layer experiences an exactly opposite proximity field, causing the Rashba phases stemming from both 1T-TaS$_{2}$ interfaces to constructively interfere and add up to a giant cumulative value of almost $\pi/2$ ($\phi_{\text{R}} \approx 88^{\circ}$), regardless of the transverse gating field direction and magnitude. This strong spin-orbit coupling leads to a notable energy splitting of the conduction bands near the $K$-point [Fig.~\ref{fig:DFT_fit}(a)]. While the spins of the hole bands are aligned mostly in the $z$ direction, the conduction-like bands show more $k$-dependent, non-zero in-plane spin components, acquiring an almost pure radial Rashba spin texture as depicted on the constant energy contour in Fig.~\ref{fig:DFT_fit}(c). Furthermore, the out-of-plane spin expectation $s_z$ [Fig.~\ref{fig:DFT_fit}(b)] showcases a rapid variation around $E - E_0 \approx 6\text{ meV}$, where the spin polarization sharply switches signs.

By flipping one of the 1T-TaS$_{2}$ layers to form the AA' stacking, we place graphene into a symmetrical potential that suppresses the proximitized Rashba SOC. The horizontal mirror plane symmetry ($\sigma_{\rm h}$) rigorously forces the effective spin-orbit fields to point strictly perpendicular to the graphene plane. Consequently, the in-plane spin expectations are identically zero [Fig.~\ref{fig:DFT_fit}(f)], and the out-of-plane spin expectations $s_z$ [Fig.~\ref{fig:DFT_fit}(e)] are perfectly quantized at $\pm 0.5 \hbar$ across the entire momentum range. This rigid orthogonal locking establishes a highly stable persistent spin texture, ensuring that the bands undergo smooth parabolic dispersion [Fig.~\ref{fig:DFT_fit}(d)] without the complex hybridization present in the AA configuration.
For both stackings, the excellent agreement between the first-principles DFT calculations (circles) and the tight-binding model fits (solid lines) confirms the reliability of our effective Hamiltonian description.

\subsection{Spin and orbital angular momentum accumulation}\label{sec:charge-spin}
For an electric bias applied to the system, represented by a field $E$, the non-equilibrium spin and orbital angular momentum accumulation, $\delta \mathcal{O}$, $\mathcal{O}\in\{s_{i},L_{i}\}$ can be calculated using the linear response theory as $\delta \mathcal{O} = \chi_{\mathcal{O}} E_x$, where $\chi$ is the response function, and $s_{i} = (\hbar/2)S_{i}$.
The response function consists of the Fermi sea $\chi^{\rm{sea}}_{\mathcal{O}}(\bm{k})$ and Fermi surface $\chi^{\rm{surf}}_{\mathcal{O}}(\bm{k})$ parts, and is integrated over the first Brillouin zone
\begin{equation}
    \delta\mathcal{O}=\frac{E}{4\pi^2}\int d^2{\bm k}[\chi_{\mathcal{O}}^{\rm surf}({\bm k})+
     \chi_{\mathcal{O}}^{\rm sea}({\bm k})].
\end{equation}
The $k$-dependent Fermi sea and Fermi surface response functions can be expressed using the Kubo formula~\cite{K56,K57} in the Smrčka-Středa formulation~\cite{SS77,CB01,BM20} as
\begin{widetext}
\begin{eqnarray}
    \chi_{\mathcal{O}}^{\rm surf}({\bm k})&=&\frac{\hbar}{\pi}\gamma^2
    \sum_{n,m} \frac{{\rm Re}\{
    \langle n{\bm k}|\mathcal{O}|m{\bm k}\rangle 
    \langle m{\bm k}|j_x|n{\bm k}\rangle \}}
    {[(\epsilon_{n{\bm k}}-E_{\rm{F}})^2+\gamma^2][(\epsilon_{m{\bm k}}-E_{\rm{F}})^2+\gamma^2]}\\
      \chi_{\mathcal{O}}^{\rm sea}({\bm k})&=&\hbar
    \sum_{n\neq m} (f_{n,{\bm k}}-f_{m,{\bm k}})
    \frac{{\rm Im}\{
    \langle n{\bm k}|\mathcal{O}|m{\bm k}\rangle 
    \langle m{\bm k}|j_x|n{\bm k}\rangle\}}
    {(\epsilon_{n{\bm k}}-\epsilon_{m{\bm k}})^2}\,, \label{eq:sea}
\end{eqnarray}
\end{widetext}
where $\epsilon_{n\bm{k}}$ are the single-particle eigenenergies of the Bloch eigenstate $\ket{n\bm{k}}$ corresponding to the Hamiltonian $H_{\kappa}$.

Orbital angular momentum (OAM) is tightly connected with spin degrees of freedom in the system with a finite SOC and contributes to the net magnetic moment as well.
The modern theory of orbital magnetization \cite{Thonhauser2011:IJMPB,Vanderbilt2018:book,Aryasetiawan2019:JPCS,Lee2026:arxiv} allows for a nonzero orbital moment accumulation even though the basis of an effective tight-binding model consists of $p_{z}$ orbitals with intrinsically zero orbital angular momentum $L_{z}$ \cite{BMG23}. 
The OAM operator is defined as $\bm{L} = -\hbar/g\mu_{\rm{B}} \bm{m}$, where $\bm{m} = -e/4(\bm{r}\times\bm{v} - \bm{v}\times\bm{r})$ is the orbital magnetic moment operator, $\mu_{B}$ is the atomic Bohr magneton and $g$ is Landé $g$-factor and its value was set to 1 in the calculations.
In the basis of eigenfunctions of the Hamiltonian \eqref{C3Model} the $L_{z}$ takes the form \cite{PGM23,BMG23,BV21}
\begin{equation}
    \bra{n\bm{k}}L_{z}\ket{m\bm{k}} = \frac{e{\rm{i}}}{2g\mu_{\rm{B}}}\bra{\partial_{\bm{k}}n\bm{k}}\times \left[ H_{\kappa} -\frac{\epsilon_{n\bm{k}} + \epsilon_{m\bm{k}}}{2} \right]\ket{\partial_{\bm{k}}m\bm{k}}\,. \label{eq:Lz_general}
\end{equation}
Contrary to the atomic center approximation, where the OAM is constructed in the basis of spherical harmonics using orbital angular momentum quantum numbers and is, in general, a diagonal matrix, the $L_{z}$ defined in Eq. \eqref{eq:Lz_general} can contain off-diagonal elements as well.
The derivative of eigenfunctions can be further simplified using $\ket{\partial_{\bm{k}}n\bm{k}} = \sum_{m\neq n} \bra{m\bm{k}}\partial_{\bm{k}}H_{\kappa}\ket{n\bm{k}}\ket{m\bm{k}}/(\epsilon_{m\bm{k}} - \epsilon_{n\bm{k}})$ \cite{BV21,Thonhauser2011:IJMPB,PGM23} as follows
\begin{widetext}
\begin{equation}
    \bra{n\bm{k}}L_{z}\ket{m\bm{k}} = \frac{{\rm{i}}e\hbar^{2}}{4g\mu_{\rm{B}}} \sum_{r\neq n,m} \left( \frac{1}{\epsilon_{r\bm{k}} - \epsilon_{n\bm{k}}} + \frac{1}{\epsilon_{r\bm{k}} - \epsilon_{m\bm{k}}}\right)  \left(\bra{n\bm{k}}v_{x}\ket{r\bm{k}}\bra{r\bm{k}}v_{y}\ket{m\bm{k}} - \bra{n\bm{k}}v_{y}\ket{r\bm{k}}\bra{r\bm{k}}v_{x}\ket{m\bm{k}} \right).
\end{equation}
\end{widetext}
In our calculations, we assume a weak disorder scattering described by the phenomenological parameter~\cite{LSK+22,BM20,FBM14,ZZF+17} $\gamma=0.1$~meV, and an electronic temperature of $k_{\rm B}T=0.01$~meV entering the Fermi-Dirac functions.
Integration of the sea and surface components was restricted only to a small integration area of about 1.4 multiple of maximal Fermi contour's diameter around the $K$ and $K'$ points in the considered energy interval. 
For discretization of the integration area, we considered step of $\Delta k = 5\times10^{-7}$~\AA$^{-1}$.

An applied electrical current $\mathbf{j} = j_x \hat{\mathbf{x}}$ drives a non-equilibrium spin density accumulation $\delta \mathbf{s}$ via the Rashba-Edelstein effect (REE)~\cite{E90,DBD14,OMR+17,GIK+17} and the unconventional Rashba-Edelstein effect (UREE)~\cite{PDR+22,LSK+22,YMK+23,OSH+23}, fundamentally dictated by the underlying Rashba phase. 
While the conventional REE generates a spin accumulation orthogonal to the applied charge current ($\delta s_y$), the UREE yields a collinear spin polarization ($\delta s_x$). 
To quantify the charge to spin conversion, we define the dimensionless parameters $\alpha_{\rm REE}$ and $\alpha_{\rm UREE}$ as:
\begin{subequations}
\begin{align}
\alpha_{\rm REE} &= \frac{e v_{\rm{F}}}{\hbar}\frac{\delta s_y}{j_x}, \label{eq:REE} \\
\alpha_{\rm UREE} &= \frac{e v_{\rm{F}}}{\hbar}\frac{\delta s_x}{j_x}, \label{eq:UREE}
\end{align}
\end{subequations}
where $\delta s_x$ ($\delta s_y$) is the current-induced non-equilibrium spin density along the $x$ ($y$) direction.
The charge to spin conversion efficiencies serve as a direct, experimentally verifiable signature of the proximity-induced SOC~\cite{GCR17,GKB+19,HSI+20,KHA+20,HKZ+21,IGH+22,CSC+22}.

\begin{figure}[htp]
    \centering
\includegraphics[width=0.9\columnwidth]{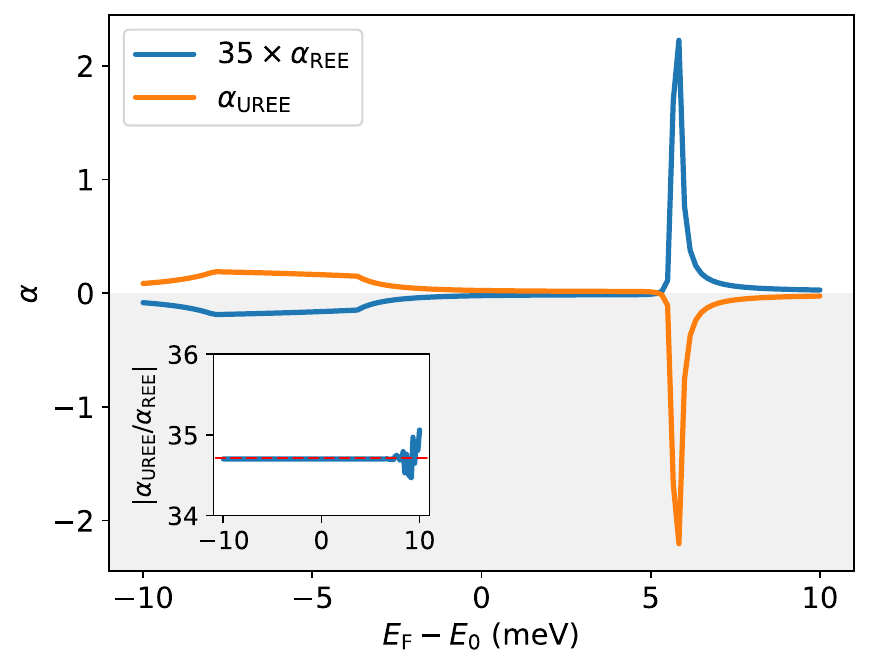}
    \caption{Charge to spin conversion coefficients in graphene encapsulated by 1T-TaS$_{2}$ in AA stacking for (un)conventional Rashba-Edelstein effect. The magnitude of $\alpha_{\rm{REE}}$ is increased by a factor of 35. The inset shows the ratio of charge to spin conversion coefficients for the unconventional and conventional cases, while the red dashed line indicates the estimated ratio.}
    \label{fig:CSC_AA}
\end{figure}

The specific $\phi_\text{R}\simeq 88^{\circ}$ of the AA stacking leads to an in-plane, radially dominated spin texture leading to a strong dominance of unconventional spin responses. As shown in Figure~\ref{fig:CSC_AA}, $\alpha_{\text{UREE}}$ exceeds $\alpha_{\text{REE}}$ by more than an order of magnitude. Crucially, the inset highlights that the ratio $|\alpha_{\text{UREE}} / \alpha_{\text{REE}}|$ maintains a remarkably flat, energy-independent plateau at a value of approximately 35 across nearly the entire spectrum. This strict proportionality is geometrically dictated by the structural phase angle, closely matching the analytical expectation ($\tan 88.35^{\circ} \approx 35$), which proves that the unconventional spin response is robustly locked by the structural design rather than details of the Fermi energy. The only fluctuations appear near $E_{\rm{F}} - E_0 \approx 6\text{ meV}$, representing a numerical sensitivity at the sharp resonance where the in-plane spin textures of the underlying conduction bands undergo inversion.

Since the system preserves the time-reversal symmetry, the net $z$-component of OAM is zero as the contribution from both $K$ valleys cancel out.
The low dimensionality of the heterostructure forbids the in-plane components of OAM as the electrons are confined to the $xy$-plane.
Although there are works that proposed a way how to calculate the in-plane OAM components for bilayer and multilayer systems \cite{HBM20,LHM23}, we neglect such a contribution to OAM for a single graphene sheet.
Therefore, no OAM is generated through the accumulation mechanism.

\subsection{Orbital and spin Hall conductivity}\label{sec:orbital-cond}
Using the expression for the sea term of the linear response function Eq.~\eqref{eq:sea}, it is possible to calculate orbital and spin Hall conductivity defined as \cite{CKR24,BMG23}
\begin{equation}
    \sigma^{\rm{OHE/SHE}} = \frac{1}{4\pi^2}\int d^2{\bm k} \,\chi_{\mathcal{O}}^{\rm sea}({\bm k})\,,
\end{equation}
where the response operator $\mathcal{O}$ is replaced by the orbital current operator $\mathfrak{J}_{y}^{z} = \{v_{y},L_{z}\}/2$ and spin current operator $\mathcal{J}_{y}^{z} = \{v_{y},s_{z}\}/2$ for OHE and SHE, respectively.
The calculations were carried on a square grid around $K$ and $K'$ points with a linear size $d = 2\times 10^{-2}$~\AA$^{-1}$ discretized to $2001\times2001$ points.
The sea term of the response function is generally not so sensitive to the sampling used in the calculation compared to the surface term.
On the other hand, a much wider area is required to converge the sea term.

\begin{figure}[htp]
    \centering
\includegraphics[width=0.9\columnwidth]{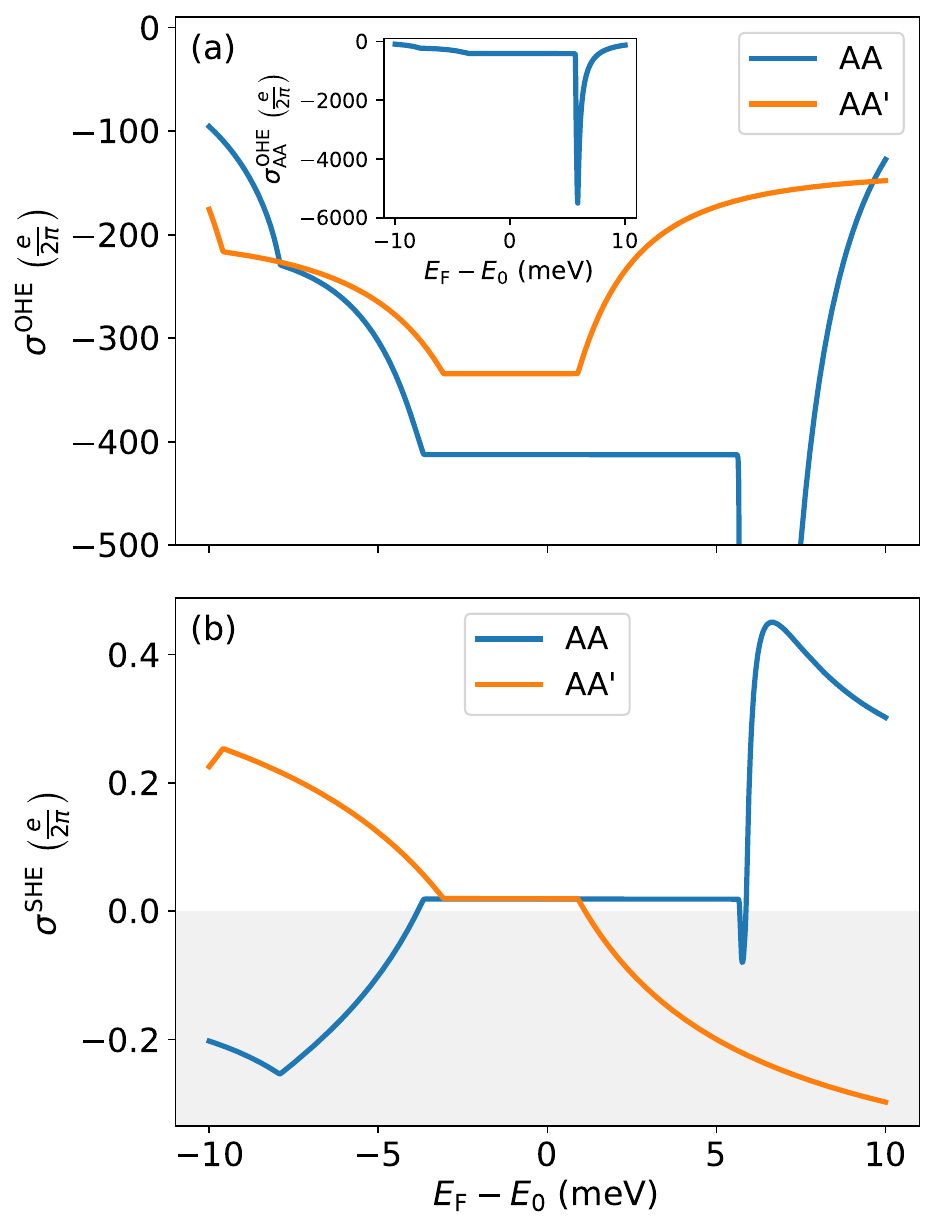}
    \caption{Calculated linear-response transport coefficients as a function of the Fermi energy $E_{\rm{F}} - E_0$ for the AA (blue lines) and AA' (orange lines) stacking configurations. 
    (a)~Orbital Hall conductivity $\sigma^{\text{OHE}}$ with plateau signatures inside the proximity-induced spectral gaps and monotonic dependences resolving discrete band edge contributions. 
    The inset highlights a pronounced resonant enhancement occurring at the energetic alignment of the two lowest conduction bands. 
    (b)~Spin Hall conductivity $\sigma^{\text{SHE}}$, displaying significantly suppressed magnitudes relative to $\sigma^{\text{OHE}}$ and exhibiting a sign inversion between the AA and AA' stackings across the charge neutrality region.}
    \label{fig:OHE}
\end{figure}

Figure~\ref{fig:OHE} shows the calculated Hall transport responses of encapsulated graphene. 
The pronounced flat plateaus extend over the spectral energy gaps opened by the proximity effects. Within these gaps, the $\sigma^{\text{SHE}}$ vanishes, indicating that the system is spin-inactive in the spin Hall regime. Conversely, the $\sigma^{\text{OHE}}$ exhibits large, finite constant values (around $-410\,(e/2\pi)$ for AA and $-335\,(e/2\pi)$ for AA' stacking).
We note that the large constant values originate from the non-local inter-site itinerant contributions, arising from the gyration of the extended Bloch wave packets in graphene in accordance with the findings reported for narrow band-gap semiconductors \cite{Pezo2022:PRB}. 
Nonetheless, the immense magnitude of $\sigma^{\text{OHE}}$, which stands roughly three orders of magnitude larger than $\sigma^{\text{SHE}}$ across all doping levels, underscores that orbital current generation remains exceptionally efficient.
Moving away from the gap, the stacking symmetry dictates the transport profiles. For the mirror-symmetric AA' stacking, the persistent spin texture leads to smooth, monotonic variations in $\sigma^{\text{SHE}}$, changing sign continuously across the neutrality point. For the AA stacking, the system displays a resonance and rapid sign inversion localized at $E_{\rm{F}} - E_0 \approx 6$~meV. This sharp transport resonance mirrors the close energetic alignment of the conduction bands that drives a divergence in the spin Berry curvature.

\section{Conclusions}\label{sec:conclusions}
In conclusion, we have systematically investigated the electronic structure, spin textures, and charge to spin/orbital transport properties of graphene encapsulated between two $\text{1T-TaS}_2$ monolayers in the CDW phase. We demonstrated that the choice of stacking configuration acts as a definitive toggle between fundamentally distinct transport and device-relevant spin regimes.
In the asymmetrical (AA) stacking, constructive interference of the proximity-induced fields yields a cumulative Rashba phase of nearly $\pi/2$. This correlation-protected radial Rashba field is robust against the application of a perpendicular electric field and leads to a significant 35-fold dominance of the unconventional Rashba-Edelstein over the conventional response across a broad energy range.
Conversely, the mirror-symmetric (AA') stacking eliminates all in-plane spin components, forcing the spins to align exclusively along the $z$-axis. While this establishes a stable persistent spin texture, the lack of in-plane spin components renders the AA' configuration less relevant for traditional charge to spin conversion architectures.
Furthermore, both configurations exhibit a giant OHE that surpasses the SHE by three orders of magnitude. Inside the proximity-induced energy gaps, the SHE vanishes while the OHE forms a rigid, finite transport plateau driven by the inter-atomic orbital angular momentum accumulation of the valence bands.
Our results showcase that the transition into the CDW phase of $\text{1T-TaS}_2$ enables fine-tuning of the proximitized spin-orbit coupling in graphene, and the stacking order enables a strategic selection between out-of-plane spin stability (AA') and efficient, correlation-protected in-plane orbital and spin current generation (AA) for future spintronic and orbitronic technologies.

\section*{Acknowledgments}
\acknowledgments
Research results were obtained using the computational resources procured in the national project National competence centre for high performance computing (project code: 311070AKF2) funded by European Regional Development Fund, EU Structural Funds Informatization of society, Operational Program Integrated Infrastructure.
J.M.~acknowledges the EU NextGenerationEU through the Recovery and Resilience Plan for Slovakia under the project No. 09I03-03-V05-00008.
M.M. acknowledges the financial support by the EU NextGenerationEU through the Recovery and Resilience Plan for Slovakia under the Project No. 09I02-03-V01-00012, by the APVV grant APVV-23-0430, and VEGA grants 2/0081/26 and 2/0133/25.
M.G.~acknowledges financial support provided by the Slovak Research and Development Agency under Contract No. APVV-SK-CZ-RD-21-0114 and by the Ministry of Education, Research, Development and Youth of the Slovak Republic, provided under Grant No. VEGA 1/0104/25 and the Slovak Academy of Sciences project IMPULZ IM-2021-42, and support of the QM4ST project funded by Programme Johannes Amos Commenius, call Excellent Research (Project No. CZ.02.01.01/00/22\_008/0004572).
\section*{References}
\bibliography{biblio}
\end{document}